\documentclass[amsmath,nofootinbib,twocolumn,superscriptaddress]{revtex4-1}
\pdfoutput=1
\usepackage[colorlinks=true,citecolor=blue,urlcolor=blue]{hyperref}
\usepackage{aas_macros,graphicx,marvosym,subfigure,amssymb,amsmath}
\usepackage{comment}
\usepackage{slashed}
\newcommand{\br}[1]{\left[#1\right]}

\newcommand{\pa}[1]{\left(#1\right)}

\newcommand{\be}{\begin{equation}}
\newcommand{\ee}{\end{equation}}
\newcommand{\bea}{\begin{eqnarray}}
\newcommand{\eea}{\end{eqnarray}}

\newcommand{\bln}{\begin{align}}
\newcommand{\eln}{\end{align}}
\newcommand{\bst}{\begin{split}}
\newcommand{\est}{\end{split}}
\newcommand{\bi}{\begin{itemize}}
\newcommand{\ei}{\end{itemize}}
\newcommand{\ben}{\begin{enumerate}}
\newcommand{\een}{\end{enumerate}}

\def\eeq{\end{equation}}

\begin{document}

\title{Black Hole Shadows, Photon Rings, and Lensing Rings}

\author{Samuel E. Gralla}
\email{sgralla@email.arizona.edu}
\affiliation{Department of Physics, University of Arizona, Tucson, Arizona 85721, USA}
\author{Daniel E.  Holz}
\email{holz@uchicago.edu}
\affiliation{Enrico Fermi Institute and Department of Physics, University of Chicago, Chicago, IL 60637, USA}
\affiliation{Department of Astronomy \& Astrophysics and Kavli Institute for Cosmological Physics, University of Chicago, Chicago, IL 60637, USA}
\author{Robert M. Wald}
\email{rmwa@uchicago.edu}
\affiliation{Enrico Fermi Institute and Department of Physics, University of Chicago, Chicago, IL 60637, USA}

\begin{abstract}

The presence of a bright ``photon ring'' surrounding a dark ``black hole shadow'' has been discussed as an important feature of the observational appearance of emission originating near a black hole.  We clarify the meaning and relevance of these heuristics with analytic calculations and numerical toy models.  The standard usage of the term  ``shadow'' describes the appearance of a black hole illuminated from all directions, including from behind the observer.  A backlit black hole casts a somewhat larger shadow.  Neither ``shadow'' heuristic is particularly relevant to understanding  the appearance of emission originating near the black hole, where the emission profile and gravitational redshift play the dominant roles in determining the observed size of the central dark area.  A photon ring results from light rays that orbit around the black hole in the near field region before escaping to infinity, where they arrive near a ring-shaped ``critical curve'' on the image plane.  Although the brightness can become arbitrarily large near this critical curve in the case of optically thin emitting matter near the black hole, we show that the enhancement is only logarithmic, and hence is of no relevance to present observations.  For optically thin emission from a geometrically thin or thick disk, photons that make only a fraction of an orbit will generically give rise to a much wider ``lensing ring,'' which is a demagnified image of the back of the disk, superimposed on top of the direct emission. For nearly face-on viewing, the lensing ring is centered at a radius $\sim 5\%$ larger than the photon ring and, depending on the details of the emission, its width is $\sim0.5$--$1M$ (where $M$ is the mass of the black hole).  It can be relatively brighter by a factor of 2--3,  as compared to the surrounding parts of the image, and thus could provide a significant feature in high resolution images.  Nevertheless, the characteristic features of the observed image are dominated by the location and properties of the emitting matter near the black hole.  We comment on the recent M87* Event Horizon Telescope observations and mass measurement.
\end{abstract}

\maketitle

\section{Introduction}

The Event Horizon Telescope (EHT) collaboration recently reported 1.3mm Very Long Baseline Interferometry (VLBI) observations of the nucleus of the nearby galaxy M87, achieving angular resolution comparable to the expected size of the supermassive black hole \cite{EHT1,EHT2,EHT3,EHT4,EHT5,EHT6}. In these papers, the concepts of a ``black hole shadow'' surrounded by a ``photon ring'' have dominated the discussion of the interpretation of the observations.  In this paper we use analytic calculations and simple models to gain better understanding and insight into the observable features of emission arising from near a black hole, paying special attention to shadows and rings.

The shadow and ring heuristics both involve a special curve on the image plane that Bardeen called the ``apparent boundary'' \cite{bardeen1973}, and that we will call the \textit{critical curve}.  By definition, when traced backwards from its observation by a distant observer, a light ray from the critical curve will asymptotically approach a bound photon orbit.  Thus, photons which are seen near the critical curve will have orbited the black hole many times on their way to the observer. For Schwarzschild, the bound orbits occur at $r=3M$ and the critical curve is a circle of apparent radius (i.e., impact parameter) $b=3\sqrt{3} M \approx 5.2M$~(see, e.g.,~\cite{luminet1979,2002ApJ...578..330H}).  For Kerr, the critical curve remains of approximately the same typical radius~\cite{bardeen1973,falcke-melia-agol2000,johanssen-psaltis2010}.

The term ``black hole shadow'' has come to represent the interior of the critical curve.  The model problem where this region corresponds to some kind of ``shadow'' is when the black hole is illuminated by a distant, uniform, isotropically emitting spherical screen surrounding the black hole (and the observer is far away from the black hole, but within the radius of the screen).  In this case the region inside the critical curve would be perfectly dark and the region outside would be uniformly bright.  As we shall discuss in subsection \ref{sec:backlit}, if a Schwarzschild black hole is instead backlit by a distant planar screen, the black hole will cast a slightly larger shadow, extending out to $b \sim 6.2M$, with a tiny amount of light emerging near the critical curve at $b\sim 5.2M$ (Fig.~\ref{fig:shadow} below).  However, neither of these shadows has much relevance for determining the appearance of emission from near a black hole, where the physical emission profile and redshift effects play the dominant roles.  For emission from an accretion disk, the main dark area extends simply to the lensed position of the inner edge of the disk~\cite{luminet1979,beckwith-done2005}.  For example, if the emission extends all the way to the horizon of a Schwarzschild black hole, the dark area (for face-on viewing) extends to $2.9M$, well within either of the ``shadows'' at $5.2M$ or $6.2M$.

The photon ring is a region of enhanced brightness near the critical curve that arises if optically thin matter emits from the region where unstable bound photon orbits exist \cite{jaroszynski-kurpiewski1997,falcke-melia-agol2000,johanssen-psaltis2010}. The light rays that comprise the photon ring can orbit many times through the emission region and thereby ``pick up'' extra brightness.  Since the optical path lengths become arbitrarily long near the critical curve, the brightness can become arbitrarily large (neglecting absorption). One of the purposes of this paper is to give a quantitative estimate of the size and observational relevance of this photon ring.

We focus primarily on the simple case of emission from an optically and geometrically thin disk near a Schwarzschild black hole, viewed face-on, since the main features can be understood from this example. We will also investigate the effects of having a geometrically thick emission region. (More detailed analysis for inclined disks in Kerr will appear separately \cite{gralla-lupsasca2019}.)  It will be convenient for us to make a distinction between a ``photon ring'' and what we call the ``lensing ring.'' We define the {\em lensing ring}\/ to consist of light rays that intersect the plane of the disk twice outside the horizon, and we define the {\em photon ring} to consist of light rays that intersect three or more times (Fig.~\ref{fig:schw} below).  For Schwarzschild, the photon ring lies at $5.19M < b < 5.23 M$ and the lensing ring lies between $5.02 M$ and $6.17 M$. For terminological definiteness, we exclude the photon ring region from the lensing ring, so that the lensing ring consists of light rays that cross the disk plane exactly twice.

Away from the lensing ring, i.e. for $b<5.02M$ or $b > 6.17 M$ in Schwarzschild, one would see only the direct emission from the disk.  Within the lensing ring, however, one would also see a lensed image of the back side of the disk superimposed upon the direct emission. This lensed image, of course, would be demagnified and distorted, and would contain varying viewing angles. Within the photon ring, one would see an additional image of the front side of the disk, and, as one gets closer to the critical curve, one would see additional alternating images of the back and front sides of the disk.

The properties of the observed emission in the photon and lensing ring regions depend significantly on the details of the emission. For example, if there is no emission at all coming from the region near the bound photon orbits, there would still be emission appearing at the photon and lensing rings arising from lensing of emission elsewhere, and there could still be some enhanced brightness effects (arising from seeing the emission region from a different ``viewing angle''), but it is not plausible that the photon and lensing rings would be prominent features of the overall emission.  As an extreme example, even if there is only emission from the disk at very large radii, there will still be a series of (highly demagnified) lensed images of the distant disk near the critical curve.  At the opposite extreme, if there is so much emission from the region near the bound photon orbits that this emission is optically thick, then there would be no photon or lensing ring effects at all---all the lensed emission would be ``blocked'' by absorption.  In intermediate cases, one could have photon and lensing rings of enhanced brightness, but their exact size and brightness would depend on the details of where the emission is coming from as well as its optical depth.  Nevertheless, we will argue in this paper that it is possible to make some general statements about the importance of the photon and lensing rings. 

\begin{figure}
    \centering
    \includegraphics[scale=.95]{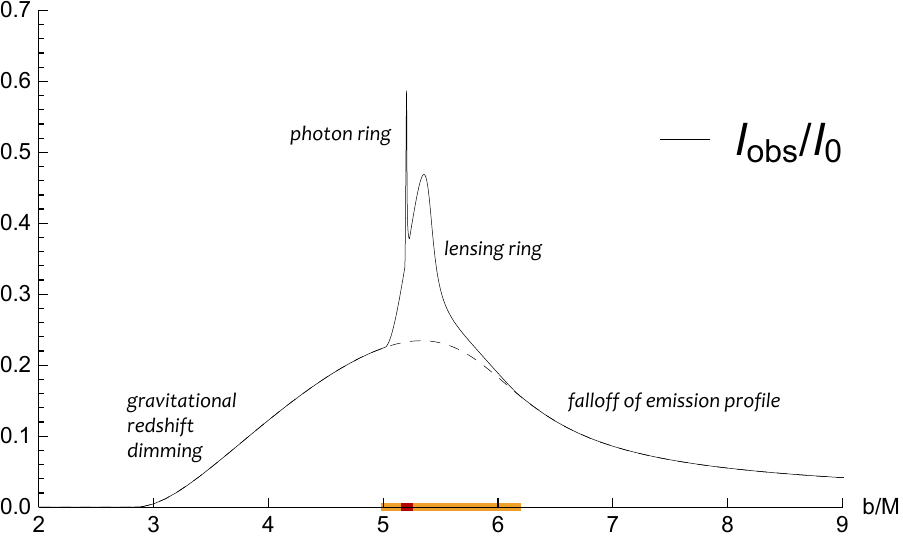}
    \caption{Using an example where the lensing ring is particularly prominent, we illustrate the key features of the image of a black hole surrounded by an optically and geometrically thin accretion disk, viewed face-on.  The observed intensity is plotted as a function of impact parameter, $b$.  This figure is an annotated version of the middle bottom plot in Fig.~\ref{fig:images} below.  The underlying emission source profile is peaked near the black hole, falling off by $r=6M$ (see bottom left panel of Fig.~\ref{fig:images}). The inner edge of the observed profile at $b \lesssim 3M$ is the lensed position of the event horizon.  (The radius of this central dark region is considerably smaller than that of the conventional ``black hole shadow'' at $r=5.2M$.) The ramp-up outside of the central dark area is due to the gravitational redshift.  The very narrow spike at $5.2M$ is the photon ring, while the distinct, broader bump at $5.4M$ is the lensing ring. The portion of this bump above the dashed line is the contribution from the image of the back side of the disk; the portion below the dashed line is from the direct emission from the front side of the disk. Beyond the lensing ring, the intensity falls off at a rate determined by the source profile. The yellow band on the $x$-axis shows the range of the lensing ring, and the red band shows the range of the photon ring.}
    \label{fig:example}
\end{figure}

A key result that allows us to make an unambiguous statement about the photon ring concerns the behavior of light rays near the critical curve.  As we have already noted above, for a Schwarzschild black hole the bound photon orbits are at $r=3M$ and the critical impact parameter is $b_c = 3\sqrt{3} M$.  We will show that in Schwarzschild, both the bending angle and the elapsed affine parameter near the black hole diverge only as $\ln |b-b_c|$ near the critical curve.  This implies that in order to additively lengthen the optical path length of a light ray through the emission region near the black hole by a given amount, one must get exponentially closer to the critical curve.  In particular, within the regime of this approximation, in order to orbit the black hole by an extra half-orbit (i.e., by an additional angle $\pi$), one must get closer to $b_c$ by a factor of $e^\pi \sim 23$.  The main consequence of this is that the light rays that make up the photon ring as defined above can contribute only a few percent of the total flux  contributed by the lensing ring.  Thus, for practical purposes the photon ring can be ignored and only the lensing ring need be considered.  We show in Appendix~\ref{sec:path} that these results extend to Kerr black holes at any inclination.

The width and brightness of the lensing ring will depend upon the geometry of the emitting region. If the emitting region is an optically thin, geometrically thin disk (viewed nearly face-on) extending near the black hole, then as we shall illustrate in section \ref{sec:thin} (see Fig. \ref{fig:schw}), the lensing ring will be narrow and should not be more than 2--3 times brighter than the direct emission (where the direct emission is included in the brightness of the lensing ring). It therefore should not make an important contribution to the observed flux at low resolution. On the other hand if the emitting region is geometrically thick and if the emission peaks near the region of bound photon orbits, then a bending angle of only $\sim \pi/2$ would be needed to significantly lengthen the optical path. The photon ring may be wide enough and bright enough to make a more significant feature in the observed total flux from the emission, as we shall illustrate in section \ref{sec:thick}. Nevertheless, even in this case, the observed emission would be dominated by the direct image of the emission profile.

The nature and properties of the enhanced emission from the photon ring and lensing ring are illustrated in Fig.~\ref{fig:example}, which corresponds to the toy model thin disk emission profile shown in the bottom left panel of Fig.~\ref{fig:images}.  The observed intensity profile is a redshifted and slightly distorted version of the source profile, upon which are superimposed lensing and photon rings. Although very bright, the photon ring is extremely narrow and makes a negligible contribution to the overall flux. The lensing ring is less bright but significantly broader, and for some emission profiles can make a modest net contribution to the total flux. The lensing ring feature is particularly prominent in this example because it lies directly on top of the broad peak of the direct emission.  The completely dark area is at $b \lesssim 3M$, far smaller than the ``shadow'' defined above.  The intensity profile shown in Fig.~\ref{fig:example} exhibits roughly the maximum possible contribution from the lensing and photon rings in the case of thin disk emission, because the emission itself is already peaked in the region of photon orbits.  For a thick disk, the lensing ring can encompass significantly more of the yellow lensing band (i.e., it can be broader), but the typical brightness enhancement will still be 2--3.  The contribution to the observed flux from the photon ring is always negligible. 

Some comparison of our terminology with that of the recent EHT papers \cite{EHT1,EHT2,EHT3,EHT4,EHT5,EHT6} is in order. In these papers, the photon ring is introduced as the precise critical curve $b=b_c$, and the associated theoretical discussion is closely tied to the unstable photon orbits.  The images arising from simulations discussed in these papers display rings of enhanced brightness, which the authors refer to as ``photon rings.''  These enhanced brightness rings cannot be ``photon rings'' as we have defined the term.  To the extent that the enhanced brightness rings are not direct features of the emission profile, they would be ``lensing rings'' in our terminology, and would therefore peak at an impact parameter roughly $5 \%$ larger than that of what we (and EHT) refer to as the ``photon ring.''

In section \ref{sec:bending} we discuss general features of photon trajectories in Schwarzschild. In section~\ref{sec:shadows} we describe the appearance of the region near a Schwarzschild black hole when it is backlit, as well as when there is emission from an optically thin, geometrically thin or thick disk near the black hole. Some comments about the case of a Kerr black hole are given in section~\ref{kerr}. In section~\ref{sec:EHT} we discuss implications of our findings for EHT observations of M87*.  In the Appendix we show that the optical path length near the critical photon orbits in Kerr is similar to the Schwarzschild case, indicating that our conclusions about photon rings will also hold for Kerr. A more complete analysis of Kerr will be given elsewhere \cite{gralla-lupsasca2019}.

\section{Light Bending Near A Schwarzschild Black Hole}
\label{sec:bending}

The behavior of null geodesics in Schwarzschild spacetime is treated in standard general relativity texts. Null geodesics possess a conserved energy $E$ and angular momentum $L$, with only their ratio $b=L/E$ relevant for the trajectory. For null geodesics that reach infinity, $b$ is the impact parameter.

\begin{figure*}
\includegraphics[scale=.9
]{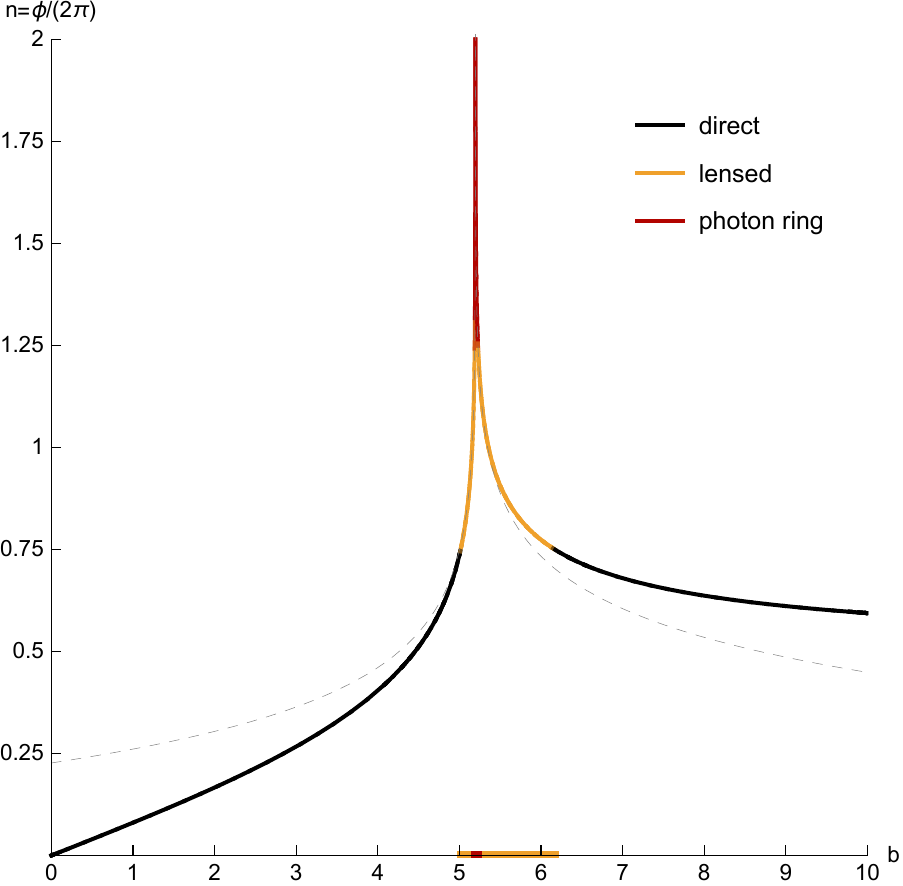}\qquad \qquad
\includegraphics[scale=.65]{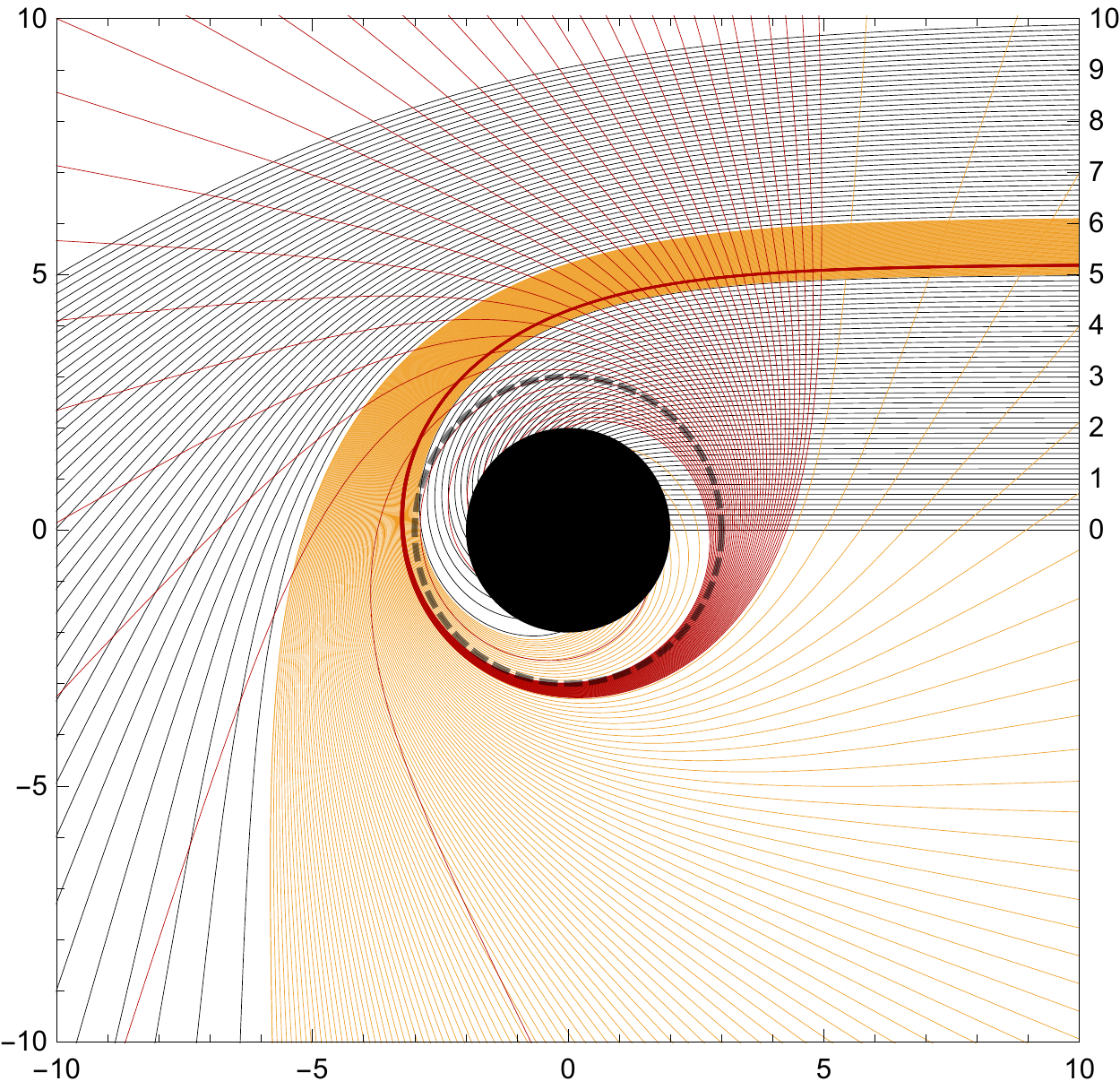}
\caption{Behavior of photons in the Schwarzschild spacetime as a function of impact parameter $b$.    On the left, we show the fractional number of orbits, $n=\phi/(2\pi)$, where $\phi$ is the total change in (orbit plane) azimuthal angle outside the horizon.  The thick line is the exact expression, while the dashed line is the approximation, Eq.~(\ref{log}).  The colors correspond to $n<0.75$ (black), $0.75<n<1.25$ (gold), and $n>1.25$ (red), defined as the direct, lensed, and photon ring trajectories, respectively.  On the right we show a selection of associated photon trajectories, treating $r,\phi$ as Euclidean polar coordinates.  The spacing in impact parameter is $1/10$, $1/100$, and $1/1000$ in the direct, lensed, and photon ring bands, respectively.  The black hole is shown as a solid disk and the photon orbit as a dashed line.
}\label{fig:schw}
\end{figure*}

We are interested in the appearance of the region near the black hole to a distant observer, for various cases of emission from near the black hole. This can be understood by tracing null geodesics from the eye of an observer backwards towards the region near the black hole. A key result needed to understand the appearance of what is observed is the total change in orbital plane azimuthal angle, $\phi$, of the null geodesic as a function of impact parameter $b$ for such trajectories. This is plotted in two ways in Fig.~\ref{fig:schw}. 

The left plot in Fig.~\ref{fig:schw} directly shows the total number of orbits, $n \equiv \phi/2 \pi$, as a function of $b$. (Note that ``straight line motion'' would correspond to $n = 1/2$, i.e., the bending of light rays that do not enter the black hole is measured by $n - 1/2$.) The most prominent feature of this plot is the singularity at $b=b_c$, where
\begin{align}
b_c=3\sqrt{3}M \approx 5.1962M.
\end{align}
Null geodesics at this critical impact parameter asymptotically approach the bound null geodesics at $r=3M$ and thus orbit around the black hole an infinite number of times.

The right plot gives perhaps a clearer picture of what an observer---placed at large distances to the right of the plot---would see.  Null geodesics emerging at impact parameter $b < 5.02 M$ originate from the black hole (or, more precisely, from the the spacetime region where the black hole was formed) and make less than $n=3/4$ orbits around the black hole. Thus, if we call the far right of the plot the ``north pole direction,'' these null geodesics cross the equatorial plane at most once. Light rays with $5.02 M < b < 6.17 M$ have $n > 3/4$ and thus cross the equatorial plane at least twice. Those with $5.19 M < b < 5. 23 M$ have $n > 5/4$ and thus cross the equatorial plane at least 3 times. Finally, the light rays with $b > 6.17$ have $n < 3/4$ and thus only cross the equatorial plane once.  For the reasons already indicated in the Introduction, we will classify these rays as follows:
\noindent
\begin{minipage}{\linewidth}
\begin{enumerate}
    \item Direct: $n<3/4$ \vspace{2mm} \\  \mbox{\qquad} $b/M\notin (5.02,6.17)$
    \item Lensed: $3/4<n<5/4$ \vspace{2mm} \\  \mbox{\qquad} $b/M\in(5.02,5.19) \textrm{ or } (5.23,6.17)$
    \item Photon ring: $n>5/4$\vspace{2mm} \\  $\mbox{\qquad} b/M\in(5.19,5.23)$
\end{enumerate}
\end{minipage}\vspace{3mm}
In Fig.~\ref{fig:schw} these are colored black, gold, and red, respectively.

It is useful to have a simple, analytic approximation to the bending angle near the critical curve $b_c$.
The Schwarzschild geodesics may be expressed in terms of elliptic integrals---see Refs.~\cite{luminet1979,ohanian1987} for details.\footnote{These references considered the case $b>b_c$ only, but the generalization is straightforward.}  Expanding the elliptic integral near $b_c$ gives the approximation,
\begin{align}\label{log}
\phi \sim \log \left( \frac{C_\pm}{|b-b_c|} \right), \qquad b \to b_c{}^\pm,
\end{align}
with
\begin{align}
C_+ &=\frac{1944}{12 + 7 \sqrt{3}} M \approx 80.6 M, \\ C_-&=648(26\sqrt{3}-45) M \approx 21.6M.
\end{align}
The result for $b>b_c$ was given by Luminet \cite{luminet1979} (see Eq.~7 of that reference).  The exact and approximate solutions are plotted in the left panel of Fig.~\ref{fig:schw}.  It can be seen that the logarithmic approximation \eqref{log} is excellent within the photon ring and most of the lensing ring, with appreciable (but still $<10\%$) deviation only at the right-most edge.

\section{Shadows and Rings}
\label{sec:shadows}

In this section we consider the appearance of a Schwarzschild black hole under various illumination conditions. In subsection \ref{sec:backlit} we consider a ``backlit'' black hole. In subsection \ref{sec:thin} we consider emission from an optically and geometrically thin disk around the black hole, viewed face-on. In subsection \ref{sec:thick} we consider emission from a geometrically thick region near the black hole.

\subsection{Backlit Black Hole}
\label{sec:backlit}

For our first model problem, consider a black hole that is illuminated from behind by a planar screen that emits isotropically with uniform brightness.  We assume that the screen is infinitely far away and infinite in extent. This problem is of no physical interest but is quite useful pedagogically for understanding features of gravitational lensing by a Schwarzschild black hole.

As in the previous section, we trace the light rays backwards from the observer.  By conservation of surface brightness (specific intensity), the image has brightness equal to that of the screen where the relevant light ray intersects the screen, and otherwise has zero brightness. The light ray will intersect the screen if and only if $b > b_c$ and $n \in (j + 1/4, j + 3/4)$ for $j=0,1,2,\dots$, where $n$ is the number of black hole orbits (see Fig.~\ref{fig:schw}). Thus, the observed brightness, $I_{\rm obs}$ is given in terms of the emitted brightness, $I_{\rm em}$ by
\begin{align}
    I_{\rm obs}(b) = \begin{cases} I_{\rm em}, & b>b_c \,\, \textrm{and} \,\, n \in(j+1/4,j+3/4) \\ 0, & \textrm{otherwise}
    \end{cases}
\end{align}
The main bright region is comprised of  the rays deflected by less than $90^\circ$, which occurs for
\begin{align}
    b>6.17M.
\end{align}
The additional bright regions have at least $n=5/4$ and are well within the validity of the log approximation \eqref{log}.  These form a sequence of rings converging to the critical curve,
\begin{align}
    \frac{b-b_c}{C_+} \in \left( e^{-2\pi(j+3/4)},  e^{-2\pi(j+1/4)} \right), \quad j=1,2,\dots
\end{align}
The widest of these rings ($j=1$) ranges from
\begin{align}
    b \in (5.1975M, 5.2274M).
\end{align}
That is, the inside edge of the first ring is already just $0.001M$  outside the critical curve and has a thickness of only $0.03M$, less than $1\%$ the critical curve radius.  The subsequent rings are exponentially closer and narrower. Thus the image features a dark hole of radius $6.17M$, together with extremely narrow rings near $5.19M$.

\begin{figure}
\includegraphics[scale=.65
]{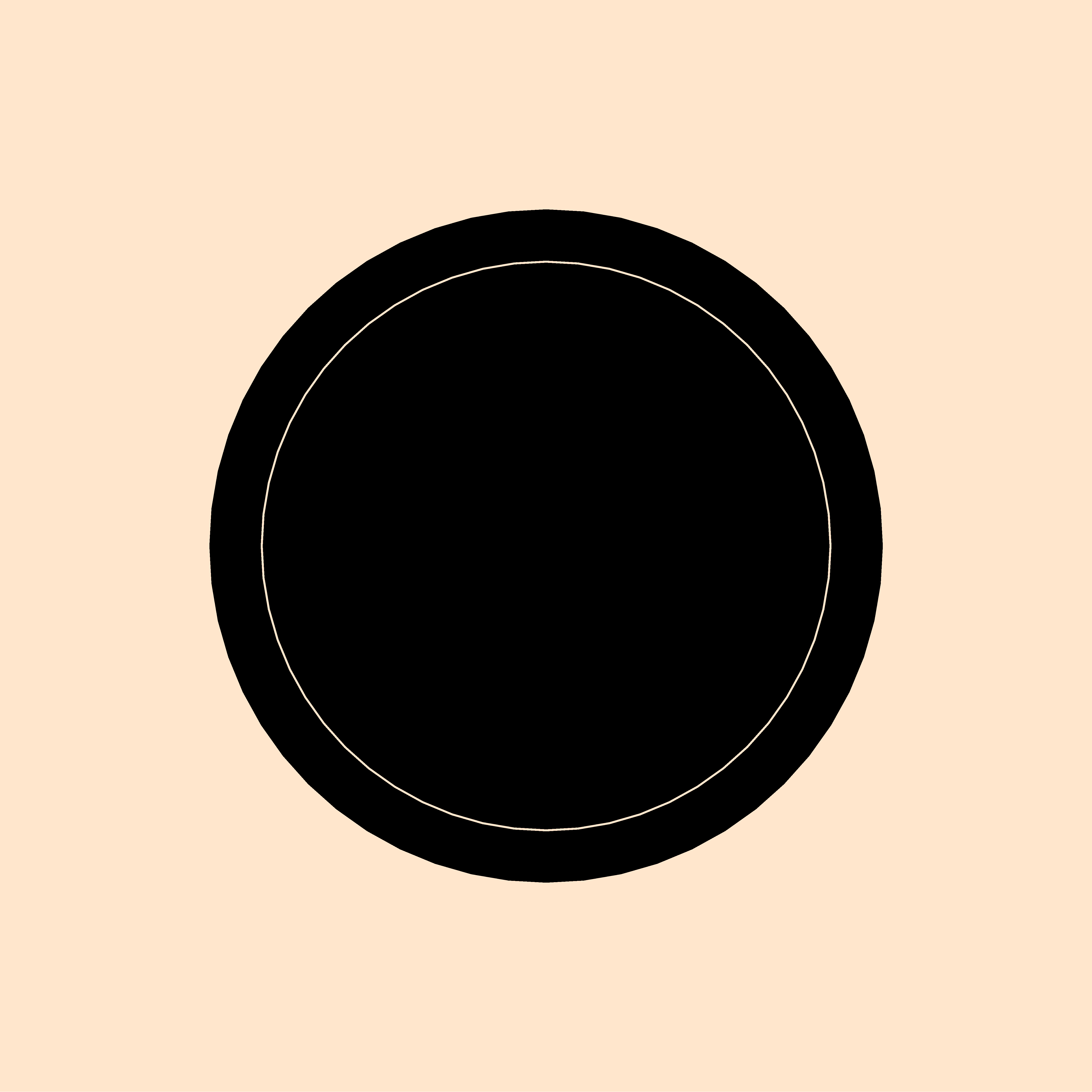}\qquad \qquad
\caption{Observational appearance of a Schwarzschild black hole that is backlit by a large, distant screen of uniform, isotropic emission.  The brightness (beige color) is uniform  where it is non-zero.  The large dark area has radius $6.17M$, and the thin ring of light has radius $5.20M$ and thickness $0.03M$. Inside of this ring is an infinite sequence of tinier and tinier rings, which are far too thin to display in this figure. We would regard the larger dark area of radius $6.17 M$ as the black hole ``shadow'', but the standard usage of this term is to refer the region inside the tiny rings as the shadow.}\label{fig:shadow}
\end{figure}

The observational appearance of a backlit black hole is shown in Fig.~\ref{fig:shadow}.  We would describe this arrangement as the black hole casting a ``shadow'' of radius $6.17M$, into which a tiny amount of light has managed to ``sneak through'' to occupy less than $1\%$ of the shadow area.  However, the standard usage of the term ``black hole shadow'' refers to the smaller portion within the thin ring.  

Finally, suppose that, instead of a distant planar screen, we had a distant spherical screen surrounding the black hole---still emitting isotropically and with uniform brightness---and a distant observer inside the radius of the screen. Then it can be seen immediately from a similar analysis that the region $b > 3 \sqrt{3} M \approx 5.20 M$ would appear uniformly bright and the region $b < 3 \sqrt{3} M $ would be entirely dark. Thus, there would be a smaller ``shadow'' and there would be no ``rings'' around the shadow.

We note in passing that Luminet \cite{luminet1979} considered illumination by a plane-parallel beam of light, mainly discussing the appearance of the deflected beam in that particular case. 

\subsection{Optically and Geometrically Thin Disk Emission}
\label{sec:thin}

We now consider some simple examples 
where the emission originates near the black hole from an optically thin and geometrically thin disk---viewed in a face-on orientation---whose specific intensity $I_\nu$ depends only on the radial coordinate.  We assume that the disk emits isotropically in the rest frame of static worldlines (i.e., the matter is at rest).
It would not be difficult to consider the much more realistic cases of orbiting and/or infalling matter, but for the face-on disk these effects are degenerate with the choice of radial profile. Our examples below are intended only as toy models, designed to illustrate the effects of gravitational lensing, as well as gravitational redshift.

We take the disk to lie in the equatorial plane, with our observer at the North pole. We denote the emitted specific intensity by 
\begin{align}
    I^{\rm em}_\nu=I(r),
\end{align}
where $\nu$ is the emission frequency in a static frame.  Since $I_\nu/\nu^3$ is conserved along a ray, radiation emitted from a radius $r$ and received at any frequency $\nu'$ has specific intensity
\begin{align}
    I^{\rm obs}_{\nu'} = g^3 I(r), \quad g=\sqrt{1-2M/r}.
\end{align}
The integrated intensity $I=\int I_\nu d\nu$ then scales as $g^4$,
\begin{align}
    I^{\rm obs} = g^4 I(r).
\end{align}

We assume that the disk is optically thin, since no interesting photon ring or lensing ring effects will occur otherwise. If a light ray followed backward from the observer intersects the disk, it will pick up brightness from the disk emission. But if the light ray has an impact parameter $b$ such that, in the notation of Fig. \ref{fig:schw}, we have $n > 3/4$, then the light ray will bend around the black hole and hit the opposite side of the disk from the back. It will therefore pick up additional brightness from this second passage through the disk. If $n > 5/4$, the light ray will also hit the front side of the disk again.  The observed intensity is a sum of the intensities from each intersection,
\begin{align}
    I^{\rm obs}(b) = \sum_m g^4 I|_{r=r_m(b)},
\end{align}
where $r_m(b)$ is the radial coordinate of the $m^{\rm th}$ intersection with the disk plane outside the horizon.  Here we have neglected absorption, which would decrease the observed intensity resulting from the additional passages.

\begin{figure}
\includegraphics[scale=.9
]{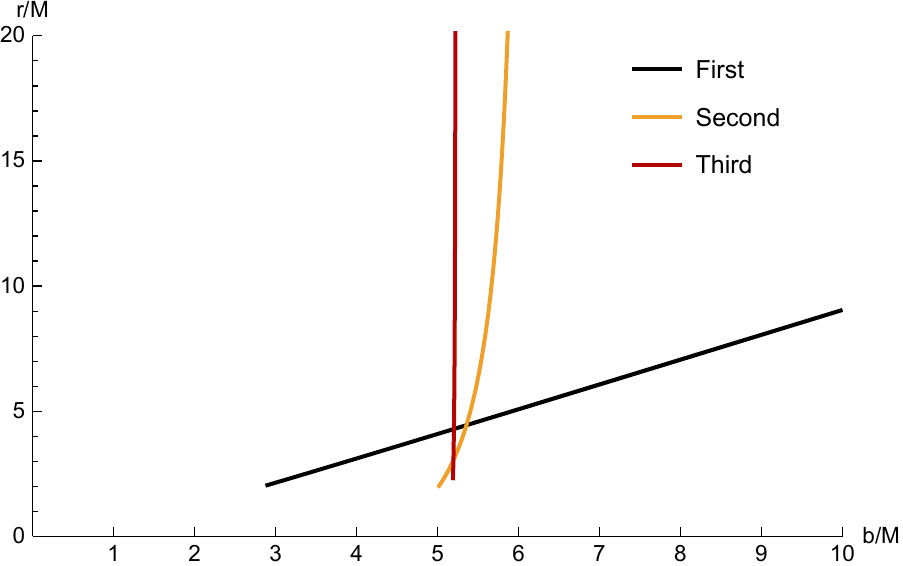}
\caption{The first three transfer functions $r_m(b)$ for a face-on thin disk in the Schwarzschild spacetime.  Tracing a photon back from the detector, these represent the radial coordinate of the first (black), second (gold), and third (red) intersections with a face-on thin disk outside the horizon.}\label{fig:transfer}
\end{figure}

We will refer to the functions $r_m(b)$ $(m=1,2,3,\dots)$ as \textit{transfer functions}. The transfer function directly shows where on the disk a light ray of impact parameter $b$ will hit. The slope of the transfer function, $dr/db$, at each $b$ yields the demagnification factor (relative to $r$, rather than proper distance) at that $b$. 
The first three transfer functions are plotted in Fig.~\ref{fig:transfer}. 
None of the transfer functions has support for $b\lesssim 2.9M$, so no light appears inside this radius. Note that this dark region is much smaller than the ``black hole shadow'' of the previous subsection. The first ($m=1$) transfer function corresponds to the ``direct image'' of the disk. The slope is nearly~$1$ over its entire range, so the direct image profile is essentially just the redshifted source profile.
The second ($m=2$) transfer function has support in the ``lensing ring'' $b/M \in (5.02,6.17)$ (including in the ``photon ring'' portion of this range of $b$).  In this range of $b$, the observer will see a highly demagnified image of the back side of the disk, with a variable demagnification given by the slope of the curve.  Over the displayed range of $r$ the average slope is around $20$, indicating that the secondary image is around 20 times smaller, and hence will typically contribute around $5\%$ of the total flux.  Finally, the third transfer function has support only in the ``photon ring'' $b/M \in (5.19,5.23)$. In this range of $b$, one will see an extremely demagnified image of the front side of the disk. This image---as well as the further images---is so demagnified that it will always contribute negligibly to the total flux.

The negligible contribution of the photon ring to the total flux can be seen analytically as follows.  Let us denote the edges of the $m^{\rm th}$ image by $b^\pm_m$,
\begin{align}
m^{\rm th} \textrm{ \ image:} \qquad b \in (b_m^-,b_m^+),
\end{align}
where $m=1,2,3^+$ corresponds to direct, lensed, and photon ring, respectively.  Inverting Eq.~\eqref{log} gives
\begin{align}\label{exp}
|b-b_c| \sim C_{\pm} e^{-\phi}, \qquad b \to b_c{}^{\pm} \, .
\end{align}
Thus, the width $\Delta b_m=b_m^+-b_m^-$ is exponentially suppressed,
\begin{align}
\Delta b_m \approx e^{-\pi} \Delta b_{m-1}, \qquad m \to \infty \, ,
\end{align}
i.e., the images are exponentially demagnified.

\begin{figure*}
\includegraphics[scale=.705]{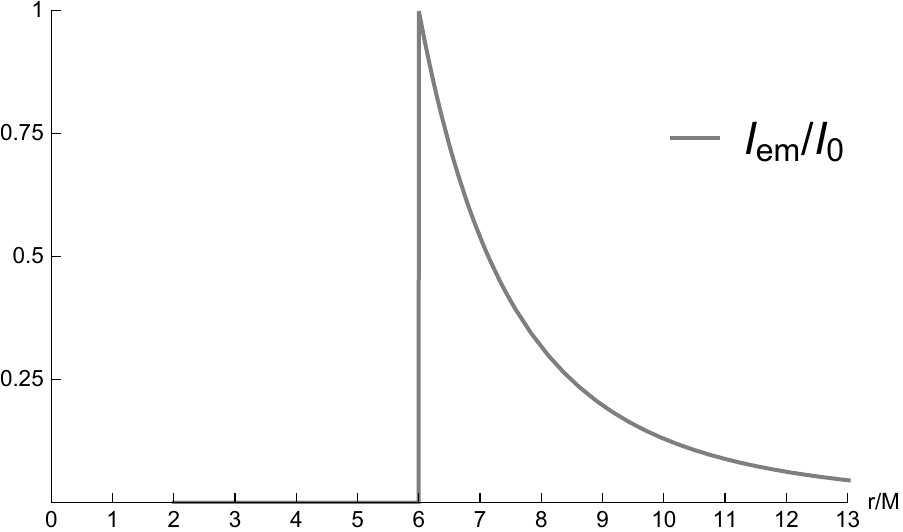} \ \ \ \includegraphics[scale=.705]{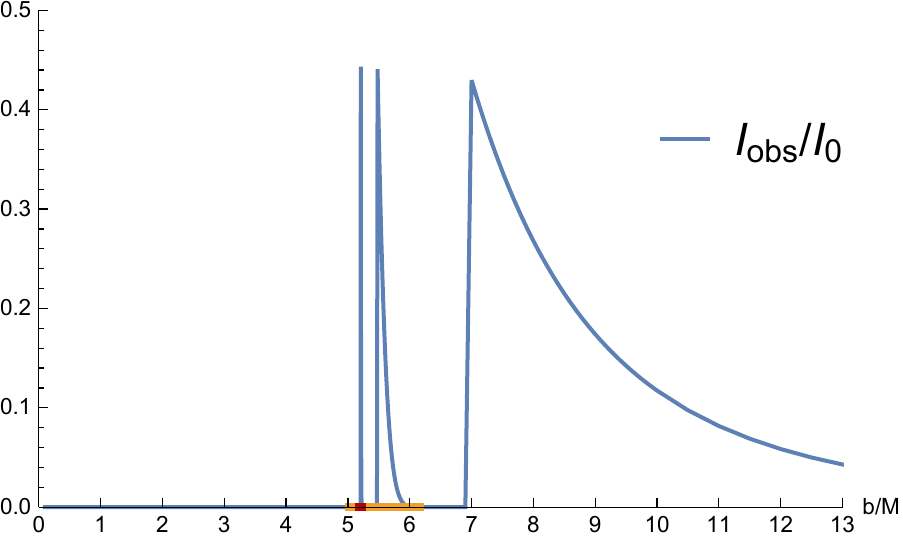} \ 
\includegraphics[scale=.097]{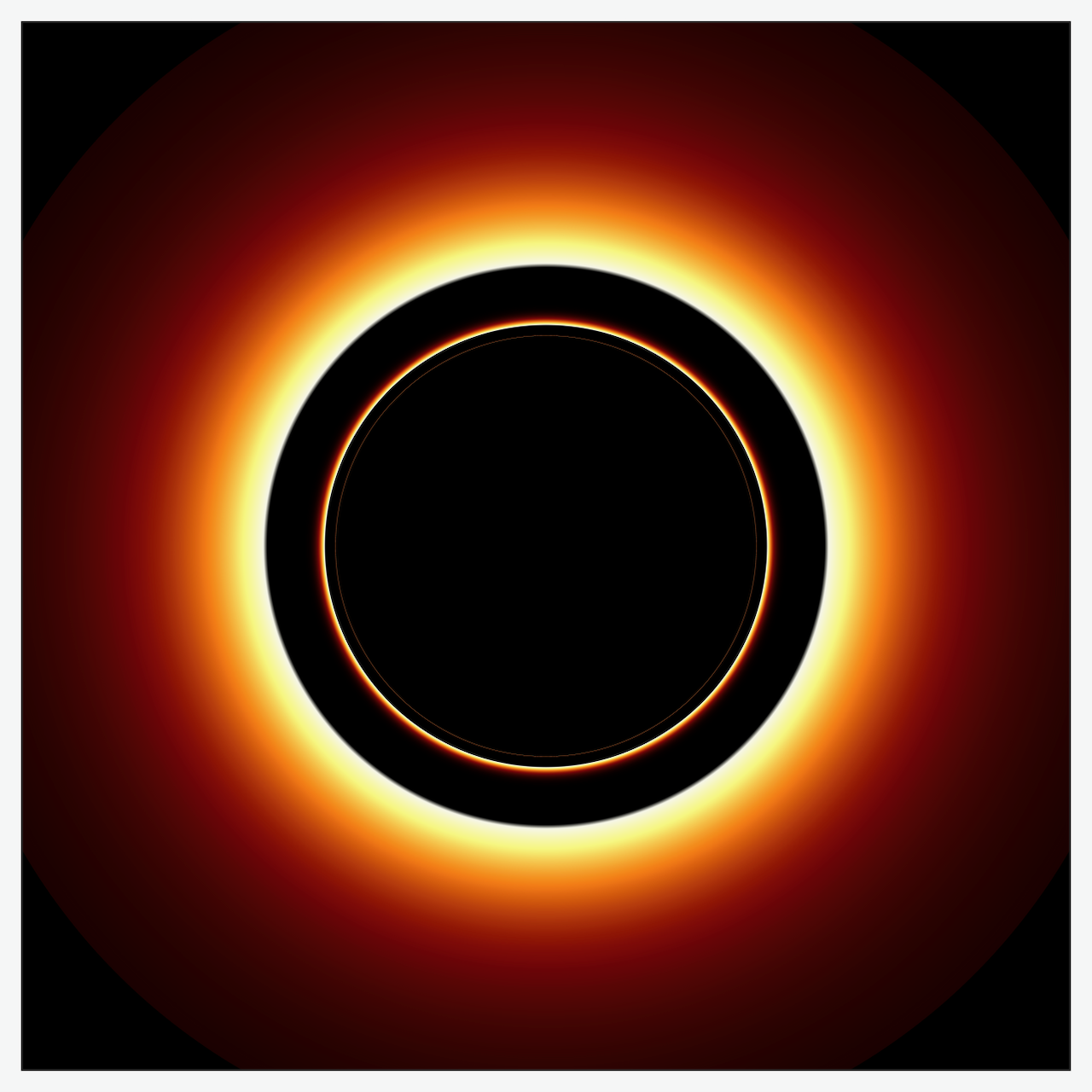} \\
\includegraphics[scale=.705]{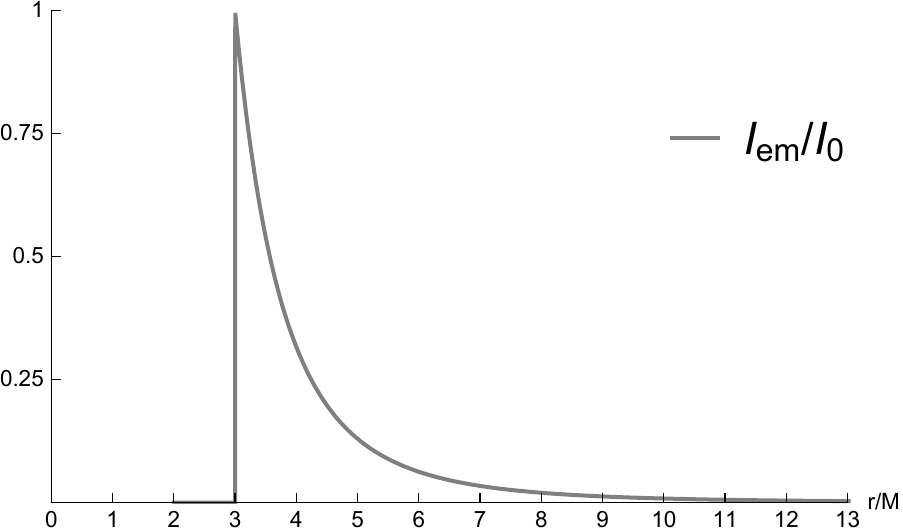} \ \ \ \includegraphics[scale=.705]{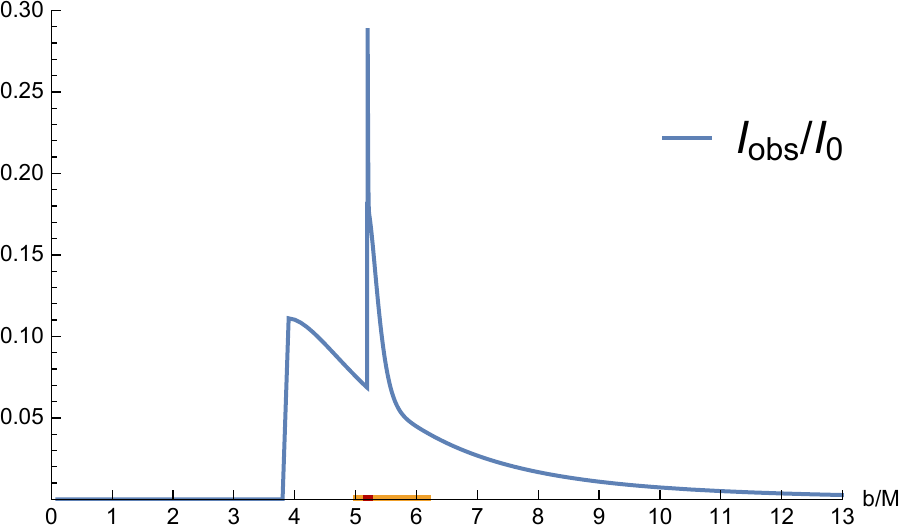} \
\includegraphics[scale=.097]{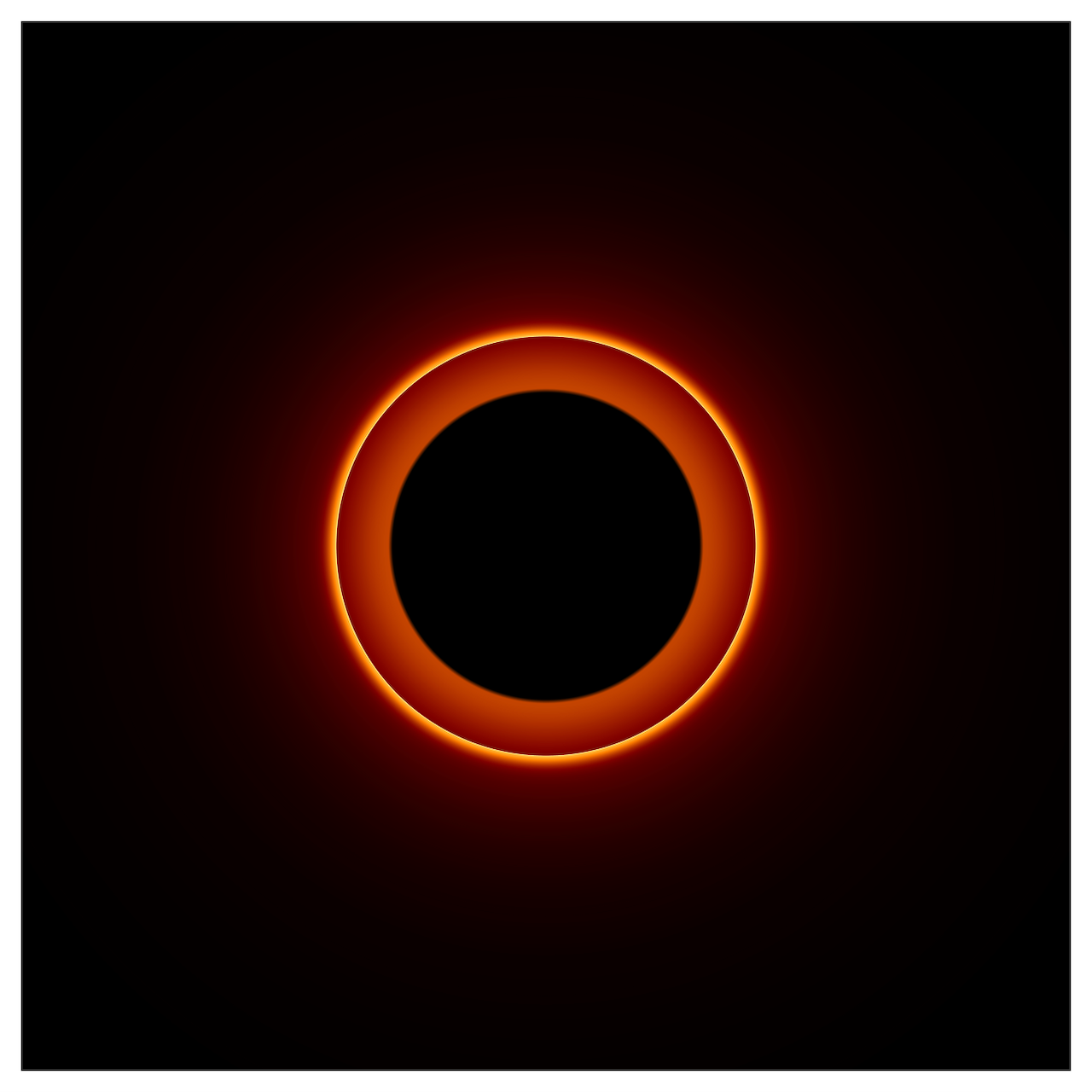} \\
\includegraphics[scale=.705]{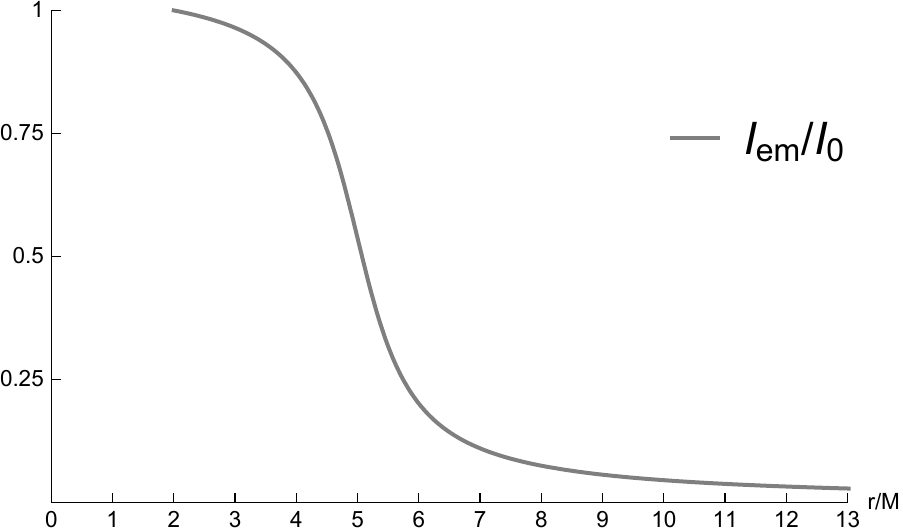} \ \ \ \includegraphics[scale=.705]{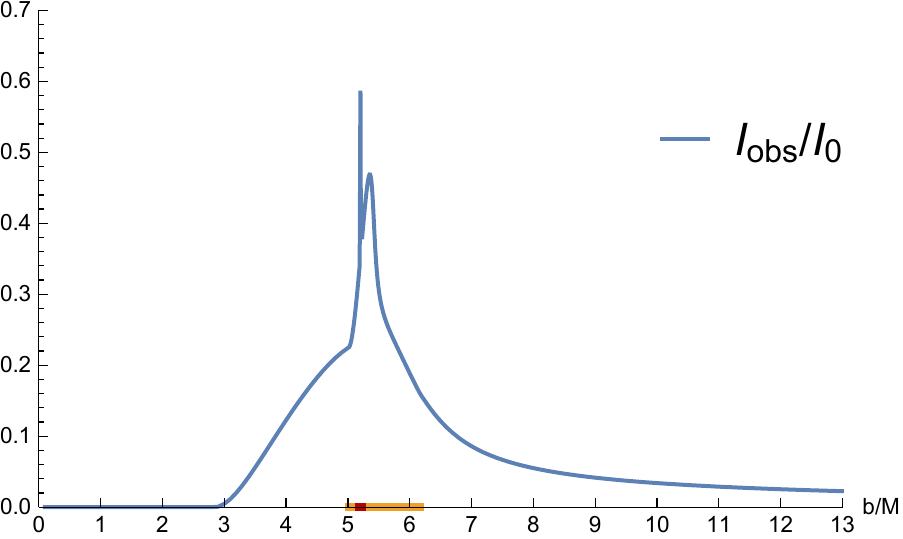} \
\includegraphics[scale=.097]{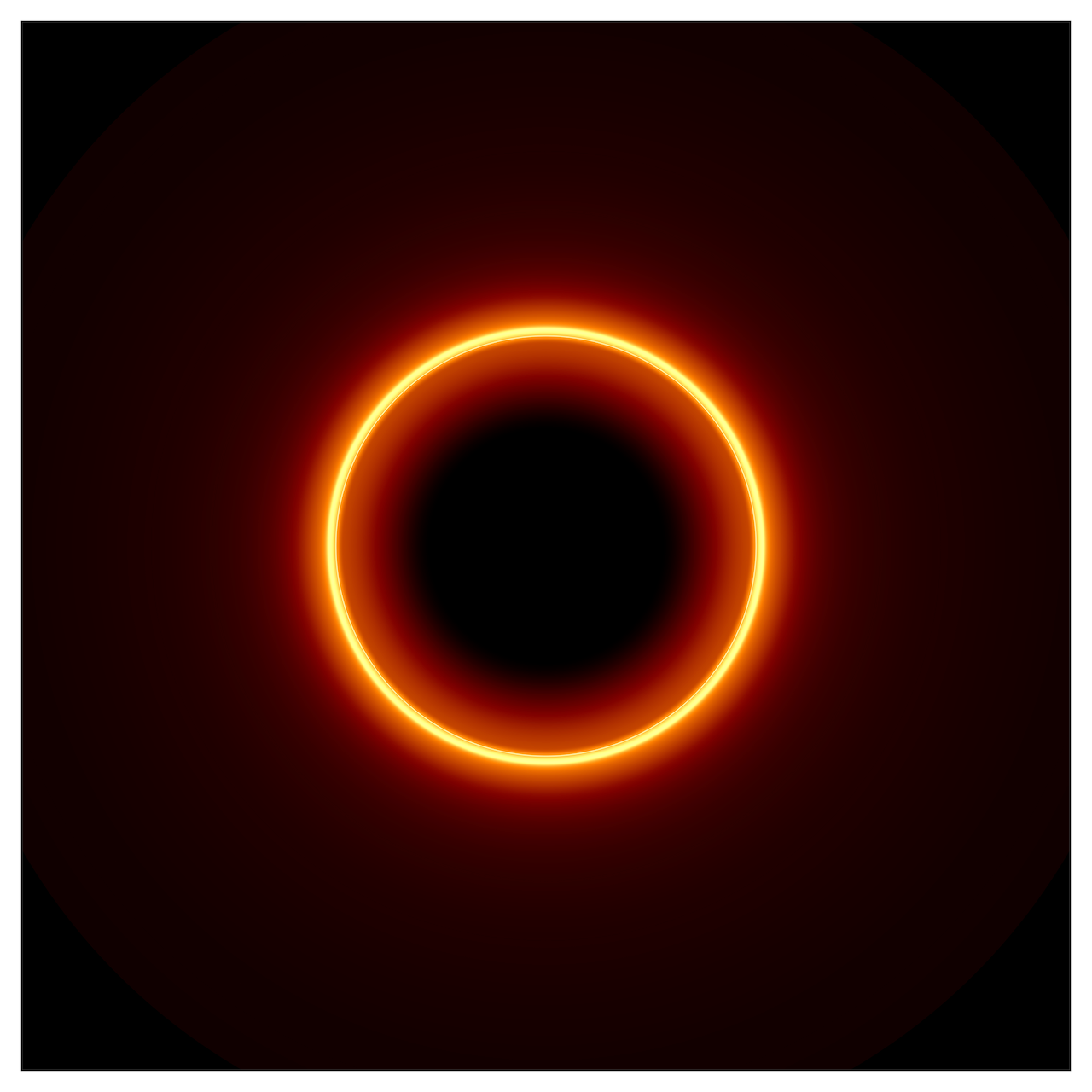} \\
\caption{Observational appearance of a thin, optically thin disk of emission near a Schwarzschild black hole, viewed from a face-on orientation.  The emitted and observed intensities $I_{\rm em}$ and $I_{\rm obs}$ are normalized to the maximum value $I_0$ of the emitted intensity outside the horizon.  The lensing ring at around $5.5M$ is clearly visible, while the photon ring at $5.2M$ is negligible.  (Only the first three images ($m=1,2,3$) are included in these plots.)  When the emission stops at some inner edge (top two rows), the radius of the main dark hole is the apparent position of the edge.  When the emission extends to the horizon, the radius of the main dark hole is the apparent position of the horizon (here $b \sim 3M$).  The critical curve $b=5.2M$ (previously called the ``shadow'') plays no role in determining the size of the main dark area.}\label{fig:images}
\end{figure*}

For optically thin emitting matter present in the region of photon orbits, the images will superpose. The local brightness can then become arbitrarily large.  However, because of the exponential demagnification, the \textit{average brightness}---which is proportional to flux in a detector---remains low for all reasonable profiles.  For example, suppose that the direct emission from near the photon orbit has typical brightness $I_{\rm local}$, while the rest of the disk has typical brightness $I_{\rm disk}$.  If $I_{\rm ring}$ is the average observed brightness in the lensing ring regime, $\sim 5M$--$6M$, we have
\begin{align}
I_{\rm ring} & \approx I_{\rm local} + I_{\rm disk}(1 + e^{-\pi} + e^{-2\pi} + \dots ). \label{sum} \\
& = I_{\rm local} + I_{\rm disk} \frac{1}{1-e^{-\pi}} \\
& \approx I_{\rm local} + 1.05 I_{\rm disk},
\end{align}
As long as the local emission is not too different from the rest of the disk (i.e., as long as there are no very bright sources outside the direct field of view), then the typical brightness enhancement is a factor of $2.05$.  The main contribution comes from the $m=2$ image (the back side of the disk), while the first ``photon ring'' image $(m=3)$ contributes just $5\%$ to the average brightness of the lensing ring.  

Fig.~\ref{fig:images} shows the appearance of the region near the black hole for a range of source profiles.  
In the example depicted in the top row, the disk emission is sharply peaked near $r=6M$, and it ends abruptly at $r=6M$ (see the left panel). Thus, in this example, the region of emission is well outside the critical photon orbits at $r=3M$. As can be readily seen from the middle panel, the direct image of the disk looks very similar to the emission profile, although its abrupt end occurs at $b \sim 7M$ due to gravitational lensing. The image of the back side of the disk (i.e., the ``lensing ring'' emission) is disjoint from the direct emission and appears in a narrow ring near $b \sim 5.5 M$. The lensing ring emission is confined to a very thin ring because the back side image is highly demagnified for $r > 6M$. As is evident from the right panel---where the lensing ring emission appears as a tiny ring inside the direct image---it makes only a very small contribution to the total flux. The photon ring emission is the extremely narrow spike at $b \sim 5.2 M$ in the middle panel of the top row. It makes a totally negligible contribution to the total flux---one can just barely see it in the right panel of the top row if one zooms in!\footnote{High resolution images will be available in the published version.}

The second row of Fig.~\ref{fig:images} depicts another example of a sharply peaked emission profile, but this time the emission peaks right at the photon orbit $r=3M$ before abruptly dropping to zero. In this case, redshift effects noticeably decrease the observed flux. However, the most important difference from the top row for our considerations is that the lensing ring and photon ring emission are now superimposed on the direct emission. This produces a lensing ring spike in the brightness from $5.2M$ to $5.5M$, with a further extremely narrow photon ring spike at $b \sim 5.2 M$. Nevertheless, the lensing ring continues to make only a very small contribution to the total flux, and the photon ring continues to make an entirely negligible contribution.

Finally, the bottom row of Fig.~\ref{fig:images} depicts emission that arises mainly from $r < 6 M$ but extends all the way down to the horizon at $r=2M$. This case was already depicted in Fig.~\ref{fig:example} above. Again, the lensing ring and photon ring are superimposed on the direct image. In this case, the lensing ring is more prominent, but the direct emission remains dominant. The photon ring continues to be entirely negligible.

Although Fig.~\ref{fig:images} shows only a few highly idealized cases of thin disk emission near a Schwarzschild black hole (viewed face on), it illustrates two key points that we believe will hold quite generally for optically thin disk emission: (1) The emission is dominated by the direct emission, with the lensing ring emission providing only a small contribution to the total flux and the photon ring providing a negligible contribution in all cases. (2) Although the photon ring always occurs near $b\sim5.2M$ and the lensing ring always occurs somewhat outside this radius, the size of the dark central area is very much dependent on the emission model.  Although the black holes in the right column of Fig.~\ref{fig:images} are the same size, the dark central areas are very different in size, ranging from radii of $b \sim 7 M$ to $b \sim 3 M$.

\subsection{Geometrically Thick Emission}
\label{sec:thick}
In this subsection, we consider emission from an optically thin but geometrically thick region near the black hole.  In this case the brightness at each impact parameter is an integrated volume emissivity along the line of sight. The observational appearance will therefore depend in a relatively complex way on both the emission profile and the shape of the emitting region.

We explore the effects of the shape of the emitting region by considering a range of highly idealized models. First, we arbitrarily select an emission region near a Schwarzschild black hole. We assign a uniform emissivity to this region and integrate the emissivity along each light ray (followed backward from the observer). To greatly simplify our calculations, instead of calculating the true optical path length of the light ray in the Schwarzschild geometry, we simply compute the Euclidean path length through the emission region (treating $r,\phi$ as polar coordinates). We also ignore redshift effects. The effects of these simplifications on our results are small compared with effects that would result from significantly varying the emission profiles in the emission region, so we do not feel that much is lost by making these simplifications. However, the reader should be aware that our aim is merely to attain a qualitative understanding of how the shape of the emission region may affect the observed appearance, not to obtain physically realistic models of geometrically thick emission.

\begin{figure*}
    \centering
    \includegraphics[scale=.45]{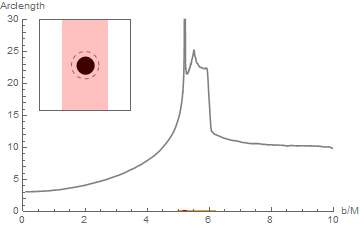}   \ \
    \includegraphics[scale=.45]{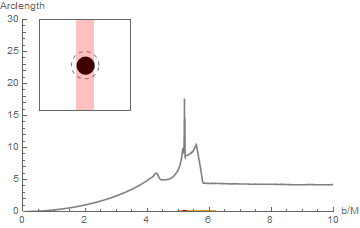}   \ \
    \includegraphics[scale=.45]{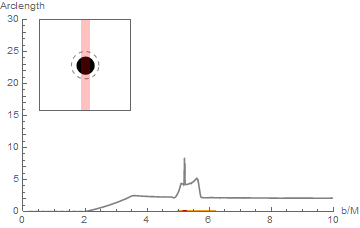}   \\
    \includegraphics[scale=.45]{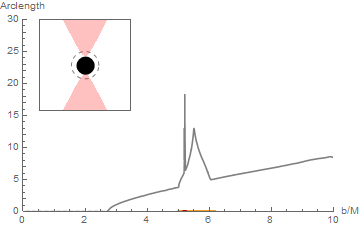}   \ \
    \includegraphics[scale=.45]{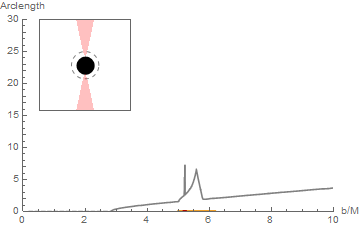}   \ \
    \includegraphics[scale=.45]{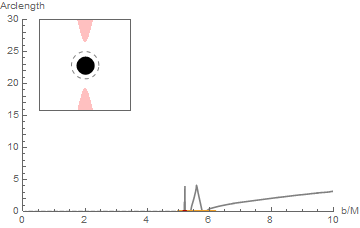}  \\ 
    \includegraphics[scale=.45]{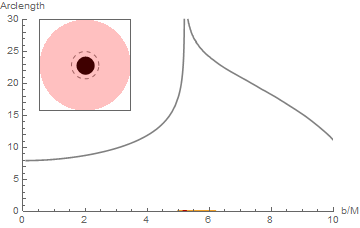}   \ \
    \includegraphics[scale=.45]{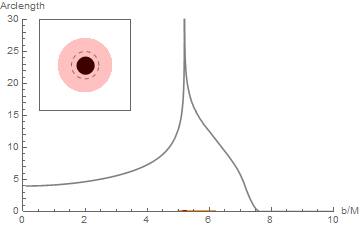}  \ \ 
    \includegraphics[scale=.45]{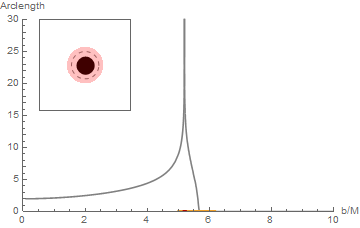}   
    \caption{Euclidean arclength of each ray with impact parameter $b$, including only the portion where it intersects a red ``emission region''.  The observer is located to the right, as in Fig.~\ref{fig:schw}.  The arclength illustrates the contribution to observed brightness from the effects of extended, optically thin emission.  For disk-like emission (top two rows) a lensing ring is clearly present, with wider disks giving wider rings.  The narrow photon ring is also visible.  (Finite numerical resolution keeps the height of the spike finite.)  When the emission is more isotropically distributed (bottom row), the lensing and photon ring features blend with the direct image.}
    \label{fig:arclen}
\end{figure*}

Fig.~\ref{fig:arclen} shows nine different choices of emission region. The top row shows plane parallel disks of various thicknesses that extend all the way to the horizon. The middle row shows tapered discs that terminate at different inner radii. The bottom row shows spherical emission regions of various radii. The corresponding observed brightness for face-on viewing (i.e., arclength)
is shown in the gray plots in the figure.

The main difference that occurs for geometrically thick emission is that a bending angle between $\pi/2$ and $\pi$ (i.e., $n$ between $3/4$ and $1$) can result in a large increase in optical path length. As a consequence, the lensing ring can be brighter and the bright region of the lensing ring can extend out to $b \sim 6 M$. Thus, the lensing ring can provide a more significant feature in the emission than in the thin disk case. However, it should be noted that the potential effects of the lensing ring are undoubtedly overemphasized in the models of the top and middle rows of Fig.~\ref{fig:arclen}, since our assumption of uniform emission over the entire volume of the emission region allows considerable brightness to be picked up from emission far from the black hole. On the other hand, for a spherical emission region, there is no clearly demarcated lensing ring feature.

The photon ring feature at $b \sim 5.2 M$ can be seen in all nine cases shown in Fig.~\ref{fig:arclen}. However, as in the thin disk case, it contributes negligibly to the total flux. This can be understood in the same manner as in the thin disk case.

In summary, for geometrically thick emission, the lensing ring can provide a more significant feature in the observed appearance than in the thin disk case. Nevertheless, the basic features of the observed appearance will be dominated by the emission profile. The photon ring contributes negligibly to the total flux.

\subsection{Inclined disks}

Thus far we have considered disks viewed from a face-on orientation.  For a disk viewed at a modest inclination (e.g., the presumed $17^\circ$ inclination of the M87* disk), we would still define the lensing and photon rings as the trajectories that make 2 and 3+ intersections with the disk plane, respectively, since this is the region of enhanced brightness.  The typical brightness enhancement will still be a factor of 2--3, with the qualitatively new feature being variation of the ring thickness around its circumference.  If $\alpha$ is a polar angle on the image plane and $\theta$ is the inclination between the line-of-sight to the observer and the disk axis ($\theta=0$ is face-on), then the first intersection with the disk plane occurs at $n=1/2+\theta/(2\pi) \sin \alpha$.  The definitions of photon and lensing rings are those given above with the replacement
\begin{align}\label{inclined}
    n \to n-\frac{\theta}{2\pi} \sin \alpha.
\end{align}
Note, however, that as $\theta \to \pi/2$ (i.e., for nearly edge-on viewing) a ``lensing ring'' defined this way would not take the shape of a ring, as the backside image of the far side of the disk makes a large contribution, comparable to that of the direct image.

In the inclined case, the brightness along the ring will also vary due to doppler-shift effects.  However, this brightness variation will precisely mirror that of the direct emission (the lensing ring is, after all, just a demagnified image of the disk), so the typical brightness enhancement will remain a factor of 2--3.  Our main conclusions are therefore unaltered in the case of 
a modestly inclined disk.

A discussion of the role of orbiting photons in the observational appearance of inclined disks was given previously by Beckwith and Done \cite{beckwith-done2005}.  Their conclusions in the nearly face-on case (Fig.~5 therein) agree qualitatively with ours.  In the nearly edge-on case, the authors claim a large contribution from orbiting photons, but the relevant photons in this case execute only a fraction of an orbit.   

\section{Kerr}
\label{kerr}

We now argue that our conclusions from the above analysis in the Schwarzschild spacetime are  qualitatively unchanged in the more realistic case of a Kerr black hole.  Our first main conclusion was that the photon ring---defined as photons that intersect the disk at least~3 times---makes a negligible contribution to the flux of an image spanning a few $M$.  The key observation is that the number of orbits increases only logarithmically with distance from the critical curve (Eq.~\eqref{log} above), making successive images exponentially demagnified.  In the appendix we prove that the affine path length diverges at most logarithmically at any point near the critical curve of the Kerr black hole.  This shows that successive images are at least exponentially demagnified, making the photon ring negligible.

Our second main conclusion from the Schwarzschild analysis is the presence of a ``lensing ring'' from the first demagnified image of an optically thin emitting disk.  Consider now a disk surrounding a Kerr black hole, lying in the plane orthogonal to the spin axis.  When the disk is viewed from a face-on angle, the main effect of the spin is to ``drag'' the photons around the viewing axis, without qualitatively affecting propagation in the radial direction.  Thus the typical properties of the lensing ring should be similar.  For nearly edge-on viewing, Kerr photons behave rather differently from Schwarzschild photons, and the properties of the lensing ring could be somewhat different.  A more complete analysis of lensing rings in the Kerr spacetime will be presented in a forthcoming paper \cite{gralla-lupsasca2019}.

\section{Implications for the Interpretation of EHT Observations of M87*}\label{sec:EHT}

The EHT collaboration has reported \cite{EHT1,EHT2,EHT3,EHT4,EHT5,EHT6} the observation of an annular feature centered on M87* with a typical radius of $\theta_{\rm obs} \sim 21 \mu\textrm{as}$.  Image reconstruction algorithms favor a width of 30--50$\%$ of the diameter, while fitting to simple ring models favors smaller fractional widths of 10--20$\%$.\footnote{A fractional width of $20\%$ corresponds to roughly one-half the nominal resolution of the array, a typical lower-bound for the scale one can probe with visibility fitting.}  The brightness along the ring is asymmetric, presumably due to doppler boosting from matter orbiting around the jet axis, which is inclined relative to the line of sight by $17^\circ$.  We will ignore the asymmetry in our discussion, instead focusing on the interpretation of the annulus.

The EHT collaboration reported a measurement \cite{EHT6} of the black hole mass that is consistent with the ``stellar dynamics'' value $6.2\times 10^9 M_\odot$~\cite{m87-mass-stellar}, but inconsistent with an alternative ``gas dynamics'' measurement of $3.5\times 10^9 M_\odot$~\cite{m87-mass-gas}. This mass measurement was made by comparing their observations with simulated images generated by applying a phenomenological prescription for electron temperature (as a function of ion density, ion internal energy, and magnetic field strength) to the results of a bank of general relativistic magnetohydrodynamic (GRMHD) simulations. These simulated images---or, at least, the ones used to fit the observations---have peak in brightness at a location $\sim10\%$ outside the photon ring.

\begin{figure}
    \centering
    \includegraphics[scale=.092]{photo2a.png}
    \includegraphics[scale=.323]{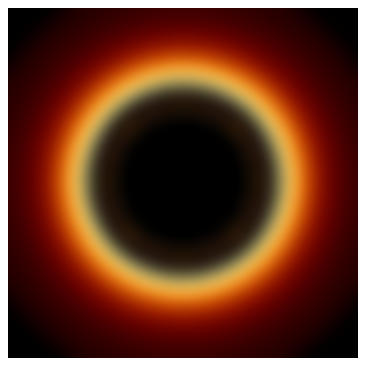} \\
    \includegraphics[scale=.092]{photo3a.png}
    \includegraphics[scale=.323]{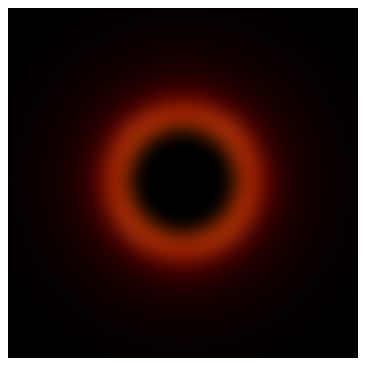} \\
    \includegraphics[scale=.092]{photo1a.png}
    \includegraphics[scale=.323]{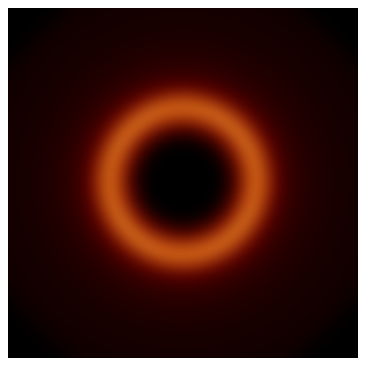}
    \caption{Images from Fig.~\ref{fig:images} before and after blurring with a Gaussian filter with standard derivation equal to $1/12$ the field of view (simulating the nominal resolution of the Event Horizon Telescope).  A lensing ring at $\sim5.4M$ appears in the sharp images, but this feature washes out after blurring.  (The photon ring at $5.2M$ is never relevant.) The effective radius of the blurred ring, and the size of the dark area, depend directly on the assumed emission profile near the black hole, with the critical curve (previously called the ``shadow'') playing no significant role.}
    \label{fig:blurry}
\end{figure}

In many places in \cite{EHT1,EHT2,EHT3,EHT4,EHT5,EHT6} it is suggested that it is a robust feature of emission models that the observed emission will peak near the photon ring.  One of the main conclusions of our analysis is that this is not the case. In this paper, we have distinguished between a ``photon ring'' (light rays that complete at least $n = 5/4$ orbits) and a ``lensing ring'' (light rays that complete between $3/4$ and $5/4$ orbits). For optically thin emission, we have argued that the ``photon ring'' {\em always} produces a sharp feature near the critical impact parameter $b_c$ ($= 3 \sqrt{3} M$ in Schwarzschild) but the peak is so narrow that it {\em never} makes a significant contribution to the observed flux. The photon ring, as we have defined it, cannot be relevant for the EHT observations. On the other hand, the lensing ring contributes a broader and more significant feature. Depending on the geometry of the emitting region and its emission profile, the lensing ring could be making a non-negligible contribution to the EHT observations. Nevertheless, the lensing ring emission is subdominant to the direct emission.

The degree of non-robustness of the observed emission peaking near the photon ring can be seen clearly if we return to the images in Fig.~\ref{fig:images} for thin disk emission and blur them to correspond roughly to the EHT resolution, as shown in  Fig.~\ref{fig:blurry}. The images of Fig.~\ref{fig:images}, of course, do not correspond to realistic emission models, but they represent a range of illustrative possibilities.  The simple blurring done in Fig.~\ref{fig:blurry} does not correspond to the EHT image reconstructions---and model fitting should of course be done in the visibility domain---but the blurring gives a rough indication of EHT's current resolution.  It can be seen from Fig.~\ref{fig:blurry} that in all cases, the blurring washes out the sharp lensing ring feature.  In the bottom row images of Fig.~\ref{fig:blurry}, the blurred image peaks at the location of the lensing ring, and the lensing ring emission itself contributes nontrivially to this peak. But in the middle row images, the peak emission is inside the lensing ring; in the top row images, the peak emission is well outside the lensing ring. Thus, it is clear that the validity of the EHT mass measurement is dependent on the validity of the detailed physical assumptions underlying the simulated images that are used to fit the observations.

\section*{Acknowledgements} It is a pleasure to acknowledge Alex Lupsasca, Dan Marrone, and Dimitrios Psaltis for helpful conversations.  SEG was supported in part by NSF grant PHY-1752809 to the University of Arizona.  DEH was supported in part by NSF grant PHY-1708081 to the University of Chicago. He was also supported by the Kavli Institute for Cosmological Physics at the University of Chicago through an endowment from the Kavli Foundation. DEH also gratefully acknowledges support from the Marion and Stuart Rice Award. RMW was supported in part by NSF grants PHY-1505124 and PHY-1804216 to the University of Chicago.

\appendix
\section{Path Length Calculation}\label{sec:path}

The key fact underlying the weakness of the photon ring effect is the slow, logarithmic increase of the path length with the distance from the critical impact parameter.  In the Schwarzschild case, photons are confined to a plane, and in Sec.~\ref{sec:bending} we discussed the number of orbits.  Photon trajectories in Kerr are more complicated, and we instead discuss the elapsed path length directly.  We first establish the general method in the Schwarzschild spacetime, before turning our attention to Kerr.

\subsection{Schwarzschild}

We use a dimensionful parameter $s$ such that $p^\mu=E dx^\mu/ds$ is the four-momentum of the photon, where $E$ is the conserved energy.  Null geodesics satisfy
\begin{align}
\frac{dr}{ds} & = \pm \sqrt{V(r)}, \qquad V=1- \frac{b^2}{r^2}\left( 1-\frac{2M}{r} \right), \label{reqn} 
\end{align}
where $r$ is the Schwarzschild coordinate, $E$ is the conserved energy and $b$ is the impact factor relative to the North pole (equal to $L/E$ in terms of the associated conserved angular momentum $L$).  The impact factor is directly proportional to the distance from the center of an image taken on the pole.  Since the Schwarzschild metric is spherically symmetric, we may place our observer on the pole without loss of generality.  Henceforth we regard $b$ as a radial coordinate on the asymptotic image.

The photon orbits have $r=3M$ and $b=3\sqrt{3} M$, and we define dimensionless fractional deviations by 
\begin{align}
r = 3M(1+\delta r), \qquad b=3\sqrt{3}M(1+\delta b).
\end{align}
To preserve the physics of interest (orbits near $3M$), $\delta b$ and $\delta r$ must be taken simultaneously to zero at the rate $\delta b \propto \delta r^2$.  Keeping the leading term in this approximation gives
\begin{align}\label{Rapprox}
V \approx 3 \delta r^2-2 \delta b.
\end{align}
Turning points occur where $V=0$, i.e. $\delta r=\pm \sqrt{(2/3)\delta b}$.  When $\delta b<0$ there are no turning points; these trajectories link the horizon and infinity.  When $\delta b>0$, the trajectories that reach infinity only involve the outermost turning point.  We therefore consider just the positive branch,
\begin{align}\label{rturn}
\delta r_{\rm turn} = \sqrt{(2/3) \delta b}.
\end{align}
We now ask how much affine parameter accumulates in a region near the photon orbit defined by
\begin{align}
-\delta R < \delta r < \delta R
\end{align}
for some positive $\delta R \ll 1$.  First consider $\delta b < 0$.  There are no turning points, and from Eq.~\eqref{reqn} we have
\begin{align}
\Delta s & = 3M \int_{-\delta R}^{\delta R} \frac{d\delta r}{\sqrt{3 \delta r^2-2 \delta b}} \\
& = \sqrt{3} M \log \left[ \frac{3}{-2\delta b}\left( \delta R+\sqrt{\delta R^2-(2/3)\delta b }\right)^2\right]. \label{woot}
\end{align}
The logarithmic divergence as $\delta b \to 0$ is clearly visible.  Now consider $\delta b>0$, where there is a single turning point.  We must also take $\delta R > \delta r_{\rm turn}$ for the photon to enter the region of interest.  Using \eqref{rturn} we then have
\begin{align}
\Delta s & = 2 \times 3M \int_{ \sqrt{(2/3)\delta b}}^{\delta R} \frac{d\delta r}{\sqrt{3 \delta r^2-2 \delta b}} \\
& = \sqrt{3} M  \log \left[ \frac{3}{2\delta b}\left( \delta R+\sqrt{\delta R^2-(2/3)\delta b }\right)^2\right].
\end{align}
Comparing with Eq.~\eqref{woot}, we see that the answer for all cases may be written
\begin{align}
\Delta s & = \sqrt{3} M  \log \left[ \frac{3}{2|\delta b|}\left( \delta R+\sqrt{\delta R^2-(2/3)\delta b }\right)^2\right] \nonumber \\ & \qquad \times \Theta\left( \delta R^2 - (2/3)\delta b\right),\label{Deltalambda}
\end{align}
where $\Theta(x)$ is zero if $x<0$ and otherwise equal to one.  It is interesting that the equation takes the same form for $\delta b > 0$ and $\delta b <0$.  This is a special property of the region $-\delta R < r < \delta R$ for Schwarzschild radius $r$ (and $\delta R \ll 1$).

For $|\delta b| \ll (3/2) \delta R^2$ we have
\begin{align}
\Delta s & \approx \sqrt{3} M\log \left( \frac{6 \delta R^2}{|\delta b|}\right). \label{Deltalambda-small}
\end{align}
This recovers the logarithmic scaling of path length with impact parameter.

It is straightforward to similarly estimate the lapse in $t$ and $\phi$.  From the definition of the conserved quantities $E$ and $b=L/E$, we have
\begin{align}
\frac{dt}{ds} & = \frac{1}{1-\frac{2M}{r}} \approx 3 \\
\frac{d\phi}{d s} & = \frac{b}{r^2} \approx \frac{1}{\sqrt{3}M}.
\end{align}
Thus from \eqref{Deltalambda-small} we have
\begin{align}
\Delta t & \approx 3\sqrt{3} M \Delta \phi, \label{Deltat} \\
\Delta \phi & \approx \log \left( \frac{6 \delta R^2}{|\delta b|}\right). 
\label{Deltaphi}
\end{align}

\subsection{Kerr}\label{sec:Kerr}

We now repeat the analysis for the non-extremal Kerr metric, using Boyer-Lindquist coordinates $(t,r,\theta,\phi)$.  We first summarize the basic results on image coordinates \cite{bardeen1973} and photon orbits \cite{teo2003} that will be needed for our analysis.  We use the notation of \cite{gralla-lupsasca-strominger2018}, except that our $\lambda$ and $q$ are their $\hat{\lambda}$ and $\hat{q}$.  The relevant derivations are also reviewed in Ref.~\cite{gralla-lupsasca-strominger2018}.  We will assume $0<a<M$, where $M$ is the mass and $a$ is the spin parameter.

Each null geodesic possesses two conserved quantities $\lambda$ and $q$, related to the angular momentum and Carter constant, respectively.  For geodesics that reach an observer at infinity at an inclination $\theta=\theta_o$, the impact parameters $(\alpha,\beta)$ are given by
\begin{align}\label{alphabeta}
	\pa{\alpha,\beta}&=\pa{-\frac{\lambda}{\sin{\theta_o}},\pm\sqrt{q^2+a^2\cos^2{\theta_o}-\lambda^2\cot^2{\theta_o}}}.
\end{align}
The $\pm$ reflects the fact that, for each $\lambda$ and $q$, there are two distinct geodesics that reach the observer.  Since $(\alpha,\beta)$ are proportional to Cartesian distance $(x,y)$ on an image, we will refer to them as ``image coordintaes''.  We emphasize that the observer is at $\theta=\theta_o$, in contrast to the Schwarzschild calculation, where one can place her on the pole without loss of generality.

Null geodesics in the Kerr metric satisfy 
\begin{subequations}
\label{eq:GeodesicEquation}
\begin{align}
	\label{eq:RadialGeodesicEquation}
	\Sigma\frac{dr}{ds}&=\pm\sqrt{\mathcal{R}(r)},\\
	\label{eq:PolarGeodesicEquation}
	\Sigma \frac{d\theta}{ds}&=\pm\sqrt{\Theta(\theta)},\\
	\label{eq:AzimuthalGeodesicEquation}
	\Sigma \frac{d\phi}{ds}&=-\pa{a-\frac{\lambda}{\sin^2{\theta}}}+\frac{a}{\Delta}\pa{r^2+a^2-a\lambda},\\
	\Sigma \frac{dt}{ds}&=-a\pa{a\sin^2{\theta}-\lambda}+\frac{r^2+a^2}{\Delta}\pa{r^2+a^2-a\lambda},
\end{align}
\end{subequations}
where $\Sigma=r^2+a^2 \cos ^2 \theta$, $\Delta=r^2+a^2-2Mr$, and
\begin{subequations}
\begin{align}
	\mathcal{R}(r)&=\pa{r^2+a^2-a\lambda}^2-\Delta\br{q^2+\pa{a-\lambda}^2},\\
	\Theta(\theta)&=q^2+a^2\cos^2{\theta}-\lambda^2\cot^2{\theta}.
\end{align}
\end{subequations}
Photon orbits at $r=\tilde{r}$ occur when  $\mathcal{R}(\tilde{r}) = \mathcal{R}'(\tilde{r})=0$.  This is possible in the range $\tilde{r}\in\br{\tilde{r}_-,\tilde{r}_+}$, where
\begin{align}
	\label{rpm}
	\tilde{r}_\pm\equiv2M\br{1+\cos\pa{\frac{2}{3}\arccos{\pm\frac{a}{M}}}},
\end{align}
and the associated conserved quantities are
\begin{subequations}\label{lambdaq}
\begin{align}
	\tilde{\lambda} &=-\frac{\tilde{r}^2\pa{\tilde{r}-3M}+a^2\pa{\tilde{r}+M}}{a\pa{\tilde{r}-M}}, \\
		\tilde{q} &=\frac{\tilde{r}^{3/2}}{a\pa{\tilde{r}-M}}\sqrt{4a^2M-\tilde{r}\pa{\tilde{r}-3M}^2}.
\end{align}
\end{subequations}
The critical curve $\{\alpha(\tilde{r}),\beta(\tilde{r})\}$ is parameterized by the radius $\tilde{r}$ of the associated photon orbit using Eqs.~\eqref{lambdaq} and \eqref{alphabeta}.  In the edge-on case $\theta_o=\pi/2$, the parameter $\tilde{r}$ ranges over the full range $\tilde{r}\in\br{\tilde{r}_-,\tilde{r}_+}$, i.e., all photon orbits are ``visible''.  In general the parameter ranges over only the subset of values such that $\beta$ is real, since only for these photon orbits can nearby photons reach infinity at the desired observation angle.

We now turn to the path length calculation.  The presence of $\Sigma(r,\theta)$ in Eqs.~\eqref{eq:GeodesicEquation} means that the $r$ and $\theta$ equations are not fully decoupled.   The path length is given by
\begin{align}
 \Delta s = \int \frac{r^2 + a^2 \cos^2 [\theta(r)]}{\pm \sqrt{\mathcal{R}(r)}} dr.
\end{align}
This expression is to be understood as an integral along a  photon path, choosing $(+)$ during portions of outward motion and $(-)$ during portions of inward motion.  Since the cosine function is positive and bounded, the path length is bounded by 
\begin{align}
\Delta s \leq \int \frac{r^2 + a^2}{\pm \sqrt{\mathcal{R}(r)}} dr.
\end{align}
For limits of integration near a photon orbit $\tilde{r}$, we have
\begin{align}\label{boogietime}
\Delta s \lesssim (\tilde{r}^2+a^2) \mathcal{I}, \qquad \mathcal{I} = \int \frac{dr}{\pm\sqrt{\mathcal{R}(r)}}.
\end{align}
It remains to compute the integral $\mathcal{I}$ for limits near a photon orbit.  Following the general approach used above for Schwarzschild, we let
\begin{align}\label{deltalambdaq}
r = \tilde{r}(1+\delta r), \quad \lambda = \tilde{\lambda} (1 + \delta \lambda), \quad q = \tilde{q}(1+\delta q),
\end{align}
and consider $\delta r \ll 1, \delta \lambda \ll 1, \delta q \ll 1$ with $\delta r^2 \sim \delta \lambda \sim \delta q$.   The radial ``potential'' is approximated as
\begin{align}
\mathcal{R} \approx C_r \delta r^2 - \delta B,
\end{align}
where
\begin{align}\label{deltaB}
\delta B = C_q \delta q + C_\lambda \delta \lambda
\end{align}
and
\begin{subequations}
\begin{align}
C_r &  = \frac{4 \tilde{r}^3}{(\tilde{r}-M)^2}\left(\tilde{r}^3 - 3 M \tilde{r}(\tilde{r} - M) -a^2 M  \right) \\
C_q & = \frac{-2\tilde{r}^3}{a^2(\tilde{r}-M)^2} \Bigg[\tilde{r}^5 - 8 M \tilde{r}^4 + \tilde{r}^3(a^2+21 M^2) \nonumber \\ & - M \tilde{r}^2 (10 a^2+ 18 M^2) +17a^2 M^2 \tilde{r} - 4 a^4 M \Bigg] \\
C_\lambda & = \frac{2 \tilde{r}^2}{a^2 (\tilde{r}-M)^2} \Bigg[ \tilde{r}^6 - 8 M \tilde{r}^5 + \tilde{r}^4(2a^2+21M^2) \nonumber \\ & - M\tilde{r}^3(10a^2+18M^2) + a^2 \tilde{r}^2(a^2 + 10 M^2)\nonumber \\ & + a^2 M \tilde{r} (6M^2-2a^2) - 3 a^4 M^2 \Bigg].
\end{align}
\end{subequations}
Turning points occur at $\delta r =\pm\sqrt{\delta B/C_r}$, provided the quantity under the square root is positive.  As before, we disregard the inner root, since a trajectory turning there will not reach infinity.  Noting that $C_r$ is positive outside the horizon, the analysis proceeds identically to the Schwarzschild case, and we find
\begin{align}\label{IIcaptain}
\mathcal{I} & = \frac{\tilde{r}}{\sqrt{C_r}} \log \left[ \frac{C_r}{|\delta B|}\left( \delta R+\sqrt{\delta R^2-\delta B/C_r }\right)^2\right] \nonumber \\ & \qquad \times \Theta\left( \delta R^2 - \delta B/C_r  \right).
\end{align}
Using Eqs. \eqref{alphabeta}, \eqref{deltalambdaq} and \eqref{deltaB}, it is straightforward to express $\delta B$ as a linear combination of the linearized deviations $\delta \alpha$ and $\delta \beta$ from the critical curve.  Then Eqs.~\eqref{IIcaptain} and \eqref{boogietime} show that the path length grows at most logarithmically as one approaches the image plane critical curve, meaning that photon ring images are at least exponentially demagnified.

\bibliographystyle{utphys}
\bibliography{photon-rings}

\end{document}